\input{aipcheck}

\documentclass[
    ,final            
  ]
  {aipproc}

\layoutstyle{6x9}

\begin{document}

\title{Density Perturbations in the Universe from Massive Vector Fields}


\author{K. Dimopoulos}{%
address={Physics Department, Lancaster University,Lancaster LA1 4YB, U.K.}}

\begin{abstract}
I discuss the possibility of using a massive vector field to generate the 
density perturbation in the Universe. I find that a scale-invariant 
superhorizon spectrum of vector field perturbations is possible to generate
during inflation. The associated curvature 
perturbation is imprinted onto the Universe following the curvaton scenario. 
The mechanism does not generate a long-range anisotropy because an oscillating 
massive vector field behaves as a pressureless isotropic fluid.
\end{abstract}

\maketitle


\section{Introduction}

The use of scalar fields in theoretical cosmology is extensive. For 
example, the dynamics of inflation is taken to be controlled by a scalar field 
(the inflaton) \cite{book}. In fact, in many inflation models several scalar 
fields are employed (e.g. hybrid inflation, assisted inflation). Alternatives 
to minimal inflation, such as the curvaton paradigm, also use scalar fields
\cite{curv}. Scalar fields have been used in other aspects of cosmology such as
Dark Energy (quintessence) or baryogenesis (Affleck-Dine mechanism). In fact 
one might wonder whether we could do any cosmology without scalar fields.

Considering that scalar fields play an important role in the Early Universe is
well motivated by the theory. Indeed, scalar fields are ubiquitous in theories 
beyond the standard model such as Supersymmetry (scalar partners, flat 
directions) and String Theory (moduli fields). However, {\em no scalar field 
has ever been observed}. This means that, designing models using unobserved 
scalar fields undermines their predictability and falsifiability, despite the 
recent precision of the observational data. 

Given the above addiction of cosmologists to scalar fields and the approaching
collider experiments which might disprove some of our favourite theories, it 
is necessary to explore other means of tackling the cosmological problems. 
Here we discuss the possibility of generating the curvature perturbation 
in the Universe using a massive vector~field.

Despite the fact that we do have evidence of the existence of both massive and
massless vector fields their use in cosmology is limited. This is so because of
the following fundamental prejudice against them: {\em Were a vector field to
dominate the Universe it would generate a large-scale anisotropy,
in conflict with the CMB observations}. A further prejudice applies to 
particle production of vector fields during inflation: {\em Particle production
of a light vector field is negligible because a massless vector field is 
conformally invariant and is not gravitationally generated during inflation}.
We will show that both these prejudices can be evaded. Moreover, there is 
evidence that some weak large-scale anisotropy might be present in the CMB 
(``Axis of Evil'' \cite{joao}).

\section{Massive Abelian Vector Fields}

\subsection{The equations of motion}

Consider a massive Abelian vector field with Lagrangian density
\begin{equation}
{\cal L}=-\frac{1}{4}F_{\rm \mu\nu}F^{\mu\nu}+\frac{1}{2}m^2A_\mu A^\mu\;,
\label{L}
\end{equation}
where, the field strength tensor is
\mbox{$F_{\mu\nu}=\partial_\mu A_\nu - \partial_\nu A_\mu$}. The equations
of motion are: \mbox{$[\partial_\mu+(\partial_\mu\ln\sqrt{-g})]
(\partial^\mu A^\nu-\partial^\nu A^\mu)+m^2A^\nu=0$},
where $g$ is the determinant of the metric tensor $g_{\mu\nu}$. We assume that
inflation lasts long enough to homogenise the Universe and inflate away its 
spatial curvature before the cosmological scales exit the horizon. Hence, we 
can use the flat-FRW metric: \mbox{$ds^2=dt^2-a^2(t)dx^idx^i$}. Similarly, we
expect inflation to homogenise the vector field too, so that 
\mbox{$A_\mu=A_\mu(t)$}. Using the above, the equations
of motion of the spatial components of the homogeneous vector field are 
\cite{vecurv}
\begin{equation}
\ddot A_i+H\dot A_i+m^2A_i=0\,,
\label{EoMhom}
\end{equation}
while for the temporal component we have \mbox{$A_t(t)=0$}, where the dot 
denotes time derivative. Perturbing the 
field equations and switching to momentum space we obtain the equations of 
motion of the Fourier components of the vector field perturbations
\cite{vecurv}
\begin{eqnarray}
&
\left[\partial_t^2+H\partial_t+m^2+\left(\frac{k}{a}\right)^2\right]
\delta{\cal A}_i^\perp 
=0 &
\label{EoMperp}\\
 & & \nonumber\\
& \hspace{-1cm} \left[\partial_t^2+\left(1+\frac{2k^2}{k^2+(am)^2}\right)
H\partial_t+m^2+\left(\frac{k}{a}\right)^2\right]
\delta{\cal A}_i^\parallel=0\,, &
\label{EoMparal}
\end{eqnarray}
where \mbox{$\delta{\cal A}_i^\parallel\equiv k_i(k_j\delta{\cal A}_j)/k^2$} is
the component parallel to $k_i$ and \mbox{$\delta{\cal A}_i^\perp\equiv
\delta{\cal A}_i-\delta{\cal A}_i^\parallel$}, with \mbox{$k^2\equiv k_ik_i$}.
Using the above equations we can investigate particle production of $A_\mu$
during inflation. Our aim is to generate a scale invariant spectrum of vector 
field perturbations. We will ignore hereafter the longitudinal component, since
$\delta{\cal A}_i^\parallel$ can generate such a spectrum only under the 
condition \mbox{$m\ll e^{-N_*}H_*$} \cite{vecurv}, where 
\mbox{$N_*\simeq 40-60$} is the number of e-folds before the end of inflation 
corresponding to the cosmological scales and $H_*$ is the inflationary Hubble 
scale. It turns out that this condition is impossible to meet if the field is 
to generate the curvature perturbation in the Universe \cite{vecurv}. Hence, 
in the following we focus on the transverse component $\delta{\cal A}_i^\perp$
and we drop the $\perp$ symbol.

\subsection{Particle production}

Solving Eq.~(\ref{EoMperp}) with boundary condition 
\mbox{$\lim_{_{_{\hspace{-.55cm}k/aH\rightarrow\infty}}}\hspace{-.5cm}
\delta{\cal A}_k=e^{ik/aH}/\sqrt{2k}$}, (i.e. matching to the Bunch-Davies 
vacuum at early times so that the perturbations start out as vacuum 
fluctuations) we get \cite{vecurv}
\begin{equation}
\delta{\cal A}_k
=\sqrt{\frac{\pi}{aH}}\;
\frac{e^{i\pi\left(\nu-\frac12\right)/2}}{1-e^{i2\pi\nu}}\;
[J_\nu(k/aH)-e^{i\pi\nu}J_{-\nu}(k/aH)],
\label{solu}
\end{equation}
where with $J_\nu$ we denote Bessel functions of the first kind and
\mbox{$\nu\equiv\sqrt{\frac{1}{4}-(m/H)^2}$}. The power spectrum of the 
generated perturbations is obtained in the limit of late times, when the
perturbations have grown to superhorizon scales. Using the above we find
\cite{vecurv}
\begin{equation}
{\cal P_A}\equiv
\frac{k^3}{2\pi^2}\hspace{-.3cm}
\lim_{_{_{\hspace{.4cm}k/aH\rightarrow 0}}}
\hspace{-.4cm}\left|\delta{\cal A}_k\right|^2
\approx\frac{8\pi|\Gamma(1-\nu)|^{-2}}{(1-\cos 2\pi\nu)}
\left(\frac{aH}{2\pi}\right)^2\left(\frac{k}{2aH}\right)^{3-2\nu}
\label{PAperp}
\end{equation}
For a light field \mbox{$\nu\simeq\frac12$} and so
\mbox{${\cal P_A}^{\rm vac}\simeq(k/2\pi)^2$}. This is the vacuum spectrum,
that can be readily obtained using the Bunch-Davies vacuum expression for 
$\delta{\cal A}_k$. Thus, in the massless limit, we recover the conformal 
invariant result where there is no particle production. However, what
we are after is a scale invariant spectrum. From Eq.~(\ref{PAperp}) it is 
evident that scale invariance requires \mbox{$\nu\simeq\frac32$}, which is 
equivalent with: \mbox{$m^2\approx -2H^2$}. 

This condition can be understood by noting that $A_\mu$ is the {\em comoving} 
vector field. Indeed, the mass term in Eq.~(\ref{L}), when using the 
flat-FRW metric, becomes 
\mbox{$\delta{\cal L}_m=\frac12m^2A_\mu A^\mu=\frac12m^2(A_t^2-a^{-2}A_iA_i)$}.
Since the Lagrangian is a physical quantity we find that the components
of the physical vector field are \mbox{$W_i\equiv A_i/a$}. Introducing these
into Eq.~(\ref{EoMhom}) during inflation we obtain:
\mbox{$\ddot W_i+3\dot W_i+(2H^2+m^2)W_i=0$} \cite{vecurv}. Hence, the equation
of motion is identical to the one of a scalar field with mass 
\mbox{$\tilde m^2=2H^2+m^2$}. Thus, a scale invariant spectrum is generated 
when the {\em physical} vector field becomes light 
\mbox{$\tilde m\rightarrow 0$}, i.e. its Compton wavelength exceeds the 
horizon. The power spectrum, in this case, is given by the Hawking temperature
\mbox{${\cal P_W}={\cal P_A}/a^2\approx(H/2\pi)^2$}.

\section{Vector Curvaton}

We now employ the scale invariant superhorizon spectrum of vector 
field perturbations obtained above to generate the curvature perturbation. 
To retain isotropy, the vector field cannot be the inflaton. The vector field 
can act as a {\em curvaton} provided it safely dominates the Universe after 
inflation. Hence, studying its post-inflationary evolution is necessary.

The stress-energy tensor is:
\mbox{$T_{\mu\nu}=\frac{1}{4}g_{\mu\nu}F_{\rho\sigma}F^{\rho\sigma}\!-
F_{\mu\rho}F_\nu^{\;\rho}\!+
m^2\left(A_\mu A_\nu-\frac{1}{2}g_{\mu\nu}A_\rho A^\rho\right)$} \cite{vecurv}.
This can be written as 
\mbox{$T_\mu^{\,\nu}={\rm diag}(\rho_A, -p_\perp, -p_\perp, +p_\perp)$}, which
is similar to a perfect fluid with the crucial difference that the pressure
along the longitudinal direction is of opposite sign to the transverse pressure
$p_\perp$, where \mbox{$\rho_A\equiv\rho_{\rm kin}+V$} and 
\mbox{$p_\perp\equiv\rho_{\rm kin}-V$}, with
\mbox{$\rho_{\rm kin}\equiv-\frac{1}{4}F_{\mu\nu}F^{\mu\nu}$} and
\mbox{$V\equiv-\frac{1}{2}m^2A_\mu A^\mu$}.
If \mbox{$p_\perp\neq 0$} then the fluid is anisotropic and cannot be allowed 
to dominate the Universe. This is why $A_\mu$ cannot be the inflaton. 

However, suppose that, after the end of inflation a phase transition renders 
the mass$^2$ positive for the vector field. This not only restores Lorentz 
invariance in the vacuum, but allows the field to begin coherent oscillations
after inflation, when \mbox{$m>H(t)$}, as evident from Eq.~(\ref{EoMhom}).
During these quasi-harmonic oscillations we have on average 
\mbox{$\overline{\rho_{\rm kin}}\approx\overline{V}$} \cite{vecurv},
i.e. \mbox{$\overline{p_\perp}\approx 0$}. Hence, the coherently oscillating 
homogeneous massive Abelian vector field behaves as pressureless 
{\em isotropic} matter. Consequently, its density decreases as 
\mbox{$\rho_A\propto a^{-3}$} \cite{vecurv}, so that it can dominate 
the radiation background without introducing significant anisotropy. Therefore,
the vector field can act as a curvaton and impose, upon domination, its own 
curvature perturbation onto the Universe. Hence, the observed curvature 
perturbation \mbox{$\zeta=5\times 10^{-5}$} can be attributed to the vector 
curvaton field, i.e.
\begin{equation}
\zeta=\zeta_A\equiv-H\frac{\delta\rho_A}{\dot{\rho}_A}=
\frac13\left.\frac{\delta\rho_A}{\rho_A}\right|_{\rm dec}
\!\simeq
\frac{2}{3}\left.\frac{\delta\bar{A}}{\bar{A}}\right|_{\rm dec}
\!=
\frac{2}{3}\left.\frac{\delta\bar{A}}{\bar{A}}\right|_{\rm osc}
\!\approx
\frac{2}{3}
\frac{a_{\rm osc}}{a_*}
\left.\frac{\delta A}{A}\right|_*\simeq\frac{H_*}{3\pi W_{\rm osc}},
\label{zA}
\end{equation}
where we defined $\zeta_A$ on a foliage of spacetime along spatially flat 
hypersurfaces and we used that: \mbox{$\rho_A\propto a^{-3}$} at the decay time
of the field (denoted by `dec');
\mbox{$\rho_A^{\rm dec}\approx 2\overline V\propto \overline{A^2}$}; 
\mbox{$\delta A/A\simeq$ const.} during oscillations but, before their onset
(denoted by `osc'), $A$ is frozen (\mbox{$A_{\rm osc}\approx A_*$}) and 
\mbox{$\delta A/A\propto a$} \cite{vecurv}; 
\mbox{$\delta A_*\simeq a_*H_*/2\pi$} and 
\mbox{$W_{\rm osc}=(A/a)_{\rm osc}\approx A_*/a_{\rm osc}$} with `*' denoting 
the epoch when the cosmological scales exit the horizon.

As shown in Eq.~(\ref{zA}), \mbox{$\zeta_A\propto\delta\rho_A/\rho_A$}.
Since $\rho_A$ is a scalar quantity, 
the generated perturbations are {\em scalar and not vector} in 
nature, despite originating from a vector field.

\section{Conclusions}

A vector field can generate the curvature perturbation in the Universe. To 
obtain a superhorizon spectrum of perturbations, its mass
should satisfy the condition: \mbox{$m^2\approx-2H^2$}. After inflation, 
the vector field can act as a curvaton provided \mbox{$m>H$} (i.e. the
vacuum is Lorentz invariant). In this case the vector field undergoes 
quasi-harmonic oscillations, during which it acts as a pressureless 
{\em isotropic} fluid. Hence, it can dominate the Universe and impose its own, 
{\em scalar} curvature perturbation without introducing any anisotropy. 

The challenge is to realise the above conditions in theories beyond the 
standard model. In supergravity, one may dispense with the requirement 
that \mbox{$m^2<0$} during inflation. Indeed, the gauge kinetic function $f$ 
of vector fields 
is expected to vary significantly during inflation. This is because 
supergravity corrections to the potential give the scalar fields masses of 
order $H_*$ \cite{eta}, which means that they are fast-rolling down the 
potential slopes resulting in \mbox{$\dot f/f\sim H$}. It can be shown that, 
if \mbox{$f\propto a^2$}, then a light vector field can obtain a scale 
invariant spectrum of perturbations without the need of \mbox{$m^2<0$}~%
\cite{sugravec}. 

\begin{theacknowledgments}
This work was supported (in part) by the European Union through the Marie Curie
Research and Training Network "UniverseNet" (MRTN-CT-2006-035863) and by STFC 
(PPARC) Grant PP/D000394/1.
\end{theacknowledgments}

\end{document}